\documentclass[aps, prd, amsmath, floats, floatfix, twocolumn, nofootinbib, superscriptaddress]{revtex4-1}
\usepackage{graphicx}
\usepackage{color}
\usepackage{latexsym}

\newcommand{\be}{\begin{eqnarray}}
\newcommand{\ee}{\end{eqnarray}}

\begin{document}

\title{Testing the Kerr nature of the supermassive black hole in Ark~564}

\author{Ashutosh~Tripathi}
\affiliation{Center for Field Theory and Particle Physics and Department of Physics, Fudan University, 200438 Shanghai, China}

\author{Sourabh~Nampalliwar}
\affiliation{Theoretical Astrophysics, Eberhard-Karls Universit\"at T\"ubingen, 72076 T\"ubingen, Germany}

\author{Askar~B.~Abdikamalov}
\affiliation{Center for Field Theory and Particle Physics and Department of Physics, Fudan University, 200438 Shanghai, China}

\author{Dimitry~Ayzenberg}
\affiliation{Center for Field Theory and Particle Physics and Department of Physics, Fudan University, 200438 Shanghai, China}

\author{Jiachen~Jiang}
\affiliation{Institute of Astronomy, University of Cambridge, Cambridge CB3 0HA, United Kingdom}

\author{Cosimo~Bambi}
\email[Corresponding author: ]{bambi@fudan.edu.cn}
\affiliation{Center for Field Theory and Particle Physics and Department of Physics, Fudan University, 200438 Shanghai, China}
\affiliation{Theoretical Astrophysics, Eberhard-Karls Universit\"at T\"ubingen, 72076 T\"ubingen, Germany}

\begin{abstract}
Einstein's theory of general relativity has been extensively tested in weak gravitational fields, mainly with experiments in the Solar System and observations of radio pulsars, and current data agree well with the theoretical predictions. Nevertheless, there are a number of scenarios beyond Einstein's gravity that have the same predictions for weak fields and present deviations only when gravity becomes strong. Here we try to test general relativity in the strong field regime. We fit the X-ray spectrum of the supermassive black hole in Ark~564 with a disk reflection model beyond Einstein's gravity, and we are able to constrain the black hole spin $a_*$ and the Johannsen deformation parameters $\alpha_{13}$ and $\alpha_{22}$ separately. For $\alpha_{22} = 0$, we find $a_* > 0.96$ and $-1.0 < \alpha_{13} < 0.2$ with a 99\%~confidence level. For $\alpha_{13} = 0$, we get $a_* > 0.96$ and $-0.1 < \alpha_{22} < 0.9$ with a 99\%~confidence level. Our measurements are thus consistent with the hypothesis that the supermassive compact object in Ark~564 can be described by the Kerr metric.
\end{abstract}

\maketitle


\section{Introduction}

Einstein's theory of general relativity has successfully passed a large number of observational tests, mainly experiments in the Solar System and accurate radio observations of binary pulsars~\cite{will}. The interest is now shifting to testing Einstein's gravity in the strong field regime, which is largely unexplored. The best laboratory for testing strong gravity is the spacetime around astrophysical black holes~\cite{rev}.

In 4-dimensional Einstein's gravity, the only stationary and asymptotically-flat vacuum black hole solution, which is regular on and outside the event horizon, is the Kerr metric~\cite{kerr1,kerr2}. The spacetime around astrophysical black holes formed by gravitational collapse is expected to be  well approximated by the Kerr geometry~\cite{book}. Nevertheless, there are a number of scenarios beyond Einstein's gravity that predict macroscopic deviations from the Kerr spacetime~\cite{new}. In particular, several authors have recently pointed out that quantum gravity may show up at the gravitational radius of a system rather than at the tiny Planck scale, suggesting the possibility of observing a signature of quantum gravity from astrophysical black holes~\cite{n1,n2,n3,n4}.

X-ray reflection spectroscopy is potentially a powerful tool for testing the strong gravity region around astrophysical black holes~\cite{k0}. In the past ten years, this technique has been developed and employed to measure black hole spins under the assumption that the spacetime metric around these objects is described by the Kerr solution~\cite{k1,k2}. Recently, the possibility of using X-ray reflection spectroscopy to test general relativity in the strong gravity regime has been explored~\cite{nk0,nk1,nk2,nk2a,nk2b,nk3,nk3b,nk4,nk5}.

{\sc relxill} is currently the most advanced X-ray reflection model to describe the reflection spectrum of thin disks around Kerr black holes~\citep{r1,r2}. Recently, we have extended this model to {\sc relxill\_nk}~\citep{noi1}. {\sc relxill\_nk} can include a variety of parametrically deformed Kerr black hole metrics. Together with the spin parameter $a_*$, these metrics are characterized by some ``deformation parameters'', which are introduced to quantify possible deviations from the Kerr spacetime.

Previous work with {\sc relxill\_nk} has provided some constraints on the Johannsen deformation parameters (defined below). In Ref.~\cite{noi2}, we analyzed some \textsl{XMM-Newton}, \textsl{NuSTAR}, and \textsl{Swift} data of the supermassive black hole in 1H0707--495 with {\sc relxill\_nk} and we constrained the deformation parameter $\alpha_{13}$. 1H0707--495 indeed shows strong reflection signatures~\cite{fabian11,kara15}, and it was thus a good source for implementing {\sc relxill\_nk}. In Ref.~\cite{Wang-Ji:2018ssh}, we analyzed the black hole binary GX339--4.

The outline of the paper is as follows. We review the Johannsen metric in Sec.~\ref{sec:metric}. The source, observation and data reduction are described in Sec.~\ref{sec:obs}. Sec.~\ref{sec:analysis} describes our data analysis with various models. Results are given in Sec.~\ref{sec:results}, followed by a discussion in Sec.~\ref{sec:discuss}.

\section{Metric\label{sec:metric}}
Presently, {\sc relxill\_nk} includes the phenomenological metric proposed by Johannsen~\cite{j-m}. In Boyer-Lindquist coordinates, the line element of the Johannsen metric with the deformation parameters $\alpha_{13}$ and $\alpha_{22}$ reads (we use units in which $G_{\rm N} = c = 1$)
\be
ds^2 &=& - \frac{\Sigma \left(\Delta - a^2 A_2^2 \sin^2\theta \right)}{B^2} \, dt^2
+ \frac{\Sigma}{\Delta} \, dr^2 + \Sigma \, d\theta^2
\nonumber\\ &&
+ \frac{\left[ \left(r^2 + a^2\right)^2 A_1^2 
- a^2 \Delta \sin^2\theta\right] 
\Sigma \sin^2\theta}{B^2} \, d\phi^2
\nonumber\\ 
&& - \frac{2 a \left[ \left(r^2 + a^2\right) A_1 A_2 - \Delta \right] 
\Sigma \sin^2\theta}{B^2} \, dt \, d\phi \, ,
\ee
where $M$ is the black hole mass, $a = J/M$, $J$ is the black hole spin angular momentum, $\Sigma = r^2 + a^2 \cos^2\theta$, $\Delta = r^2 - 2 M r + a^2$, and
\be
&& A_1 = 1 + \alpha_{13} \left(\frac{M}{r}\right)^3 \, , \quad
A_2 = 1 + \alpha_{22} \left(\frac{M}{r}\right)^2 \, , \nonumber\\
&& B = \left(r^2 + a^2\right) A_1 - a^2 A_2 \sin^2\theta \, .
\ee
The Kerr metric is recovered when $\alpha_{13} = \alpha_{22} = 0$. Note that the Johannsen metric is not expected to be ``physical", it is not a solution of Einstein's theory or any other known theory of gravity, rather it serves as a way to quantify deviations from the Kerr metric and includes several alternative theory black holes as special cases.

Following Ref.~\cite{j-m}, we can exclude a violation of Lorentzian signature or the existence of closed time-like curves in the exterior region imposing, respectively, that the metric determinant is always negative and that $g_{\phi\phi}$ is never negative for radii larger than the radius of the event horizon. This leads to the following restrictions to the values of $\alpha_{13}$ and $\alpha_{22}$
\be\label{eq-bound}
\alpha_{13} &>& - \left( 1 + \sqrt{1 - a^2_*} \right)^3 \, , \\
\label{eq-boundbis}
\alpha_{22} &>& - \left( 1 + \sqrt{1 - a^2_*} \right)^2 \, ,
\ee
where $a_* = a/M$ is the dimensionless spin parameter. The Johannsen metric is also singular when $B = 0$. Imposing that this never happens for radii larger than the radius of the event horizon, for $\alpha_{22} = 0$ we find the following constraint on $\alpha_{13}$
\be\label{eq-bound2}
\alpha_{13} > - \frac{1}{2} \left( 1 + \sqrt{1 - a^2_*} \right)^4 \, ,
\ee
and for $\alpha_{13} = 0$ we have the following constraint on $\alpha_{22}$
\be\label{eq-bound3}
\alpha_{22} < \frac{\left( 1 + \sqrt{1 - a^2_*} \right)^4}{a_*^2} \, .
\ee
Note that Eq.~(\ref{eq-bound2}) is a stronger constraint on $\alpha_{13}$ than the bound in Eq.~(\ref{eq-bound}), while Eq.~(\ref{eq-bound3}) gives an upper bound. Our reflection model {\sc relxill\_nk} covers the parameter region satisfying the constraints in Eqs.~(\ref{eq-boundbis}), (\ref{eq-bound2}), and (\ref{eq-bound3}).

\section{Observations\label{sec:obs}}
In this paper, we report our analysis of the \textsl{Suzaku} observation of the supermassive black hole in Ark~564 of~\cite{walton}. Ark~564 is classified as a narrow line Seyfert~1 galaxy at redshift $z=0.0247$. It is a very bright source in the soft X-ray band, so it has been studied by several authors. For instance, recent studies on its reflection spectrum are reported in~\cite{walton,kara13,giustini,kara17}. This source looks suitable for tests of general relativity for the following reasons. First, previous studies have shown that the inner edge of the disk may be very close to the central object, which maximizes the signatures of the strong gravity region~\cite{walton}. Second, the source has a simple spectrum. There is no obvious intrinsic absorption to complicate the determination of the reflected emission.

\textsl{Suzaku} observed Ark~564 on 26-28 June 2007 (Obs. ID~702117010) for about 80~ks. For low energies ($< 10$~keV), \textsl{Suzaku} has four co-aligned telescopes used to collect photons onto its CCD detectors X-ray Imaging Spectrometer (XIS)~\cite{koyama}. XIS is comprised of four detectors; XIS0, XIS2, and XIS3 are front-illuminated and XIS1 is back-illuminated. We only used data from the front-illuminated chips because XIS1 has a lower effective area at 6 KeV and a higher background at higher energies. XIS2 data were not used in our analysis because of the anomaly after 9~November~2006.

We used HEASOFT version 6.22 and CALDB version 20180312 for the data reduction. The raw data were reduced to screened products using AEPIPELINE script of the HEASOFT package. The XIS data were screened with the standard criterion using the ftool XSELECT~\cite{koyama}. The XIS source was extracted from the 3.5~arc minutes radius centered at the source. The background region of the same size is taken from a region as far as possible from the source to avoid contamination from the latter. Response files and area files were generated using the scripts XISRMFGEN and XISSIMARFEN, respectively. Last, the data from XIS0 and XIS3 were combined into a single spectrun using ADDASCASPEC. The data were grouped to have a minimum of 30 counts per bin in order to use $\chi^2$ statistics in our spectral analysis. We excluded the energy range 1.7-2.5~keV because of calibration issues.

\begin{table*}
\centering
\vspace{0.8cm}
\begin{tabular}{lcccccc}
\hline\hline
Model & $a$ & $b$ & $c$ & $d$ & $e$ \\
\hline
{\sc tbabs} &&&& \\
$N_{\rm H} / 10^{20}$ cm$^{-2}$ & 6.74$^\star$ & 6.74$^\star$ & 6.74$^\star$ & 6.74$^\star$ & 6.74$^\star$ \\
\hline
{\sc zpowerlaw} &&&&& \\
$\Gamma$ & $2.768^{+0.003}_{-0.003}$ & -- & -- & -- & -- \\
$z$ & 0.0247$^\star$ & -- & -- & -- & -- \\
\hline
{\sc relxill\_nk} &&&&& \\
$q$ & -- & $6.447_{-0.050}^{+0.013}$ & $6.32_{-0.27}^{+0.08}$ & $6.812_{-0.913}^{+0.025}$ & $5.87_{-0.19}^{+0.12}$ \\
$i$ [deg] & -- & $<11$ & $<20$ & $<60$ & $<22$ \\
$a_*$ & -- & $0.996_{-0.006}^{+0.001}$ & $0.985_{-0.008}^{+0.003}$ & $>0.988$ &$0.979_{-0.005}^{+0.006}$ \\
$\alpha_{13}$ & -- & $-0.65^{+0.15}$ & $-0.9_{-0.2}^{+0.2}$ & $-0.2_{-0.2}^{+0.3}$ & $-1.0_{-0.2}^{+1.0}$ \\
$z$ & -- & 0.0247$^\star$ & 0.0247$^\star$ & 0.0247$^\star$ & 0.0247$^\star$ \\
$\log\xi$ & -- & $3.090_{-0.008}^{+0.007}$ & $4.56_{-0.0}^{+0.13}$ & $2.71_{-0.24}^{+0.08}$ & $2.726_{-0.009}^{+0.072}$ \\
$A_{\rm Fe}$ & -- & $0.519_{-0.019}^{+0.0}$ & $0.53_{-0.0}^{+0.03}$ & $0.85_{-0.08}^{+0.06}$ & $0.61_{-0.11}^{+0.0}$ \\
$R$ & -- & $-1$ & $-1$ & $-1$ & $-1$ \\
\hline
{\sc relxill\_nk} &&&&& \\
$\log\xi'$ & -- & -- & $3.00_{-0.69}^{+0.06}$ & -- & $1.30_{-0.47}^{+0.16}$ \\
\hline
{\sc xillver} &&&&& \\
$\log\xi''$ & -- & -- & -- & $4.34_{-0.0}^{+0.23}$ & $4.56_{-0.0}^{+0.16}$ \\ 
\hline\hline
$\chi^2$/dof & \hspace{0.2cm} 4356.16/1403 \hspace{0.2cm} & \hspace{0.2cm} 1547.91/1397 \hspace{0.2cm} & \hspace{0.2cm} 1544.85/1395 \hspace{0.2cm} & \hspace{0.2cm} 1474.30/1395 \hspace{0.2cm} & \hspace{0.2cm} 1469.88/1393 \hspace{0.2cm} \\
& = 3.105 & = 1.108 & = 1.107 & = 1.057 & = 1.055 \\
\hline\hline
\end{tabular}
\caption{Summary of the best-fit values for the spectral models $a$ to $e$ assuming $\alpha_{13}$ free and $\alpha_{22}=0$. The reported uncertainty corresponds to the 90\% confidence level for one relevant parameter. $^\star$ indicates that the parameter is frozen to the value obtained from independent measurements. \label{t-fit}}
\end{table*}

\section{Spectral analysis\label{sec:analysis}}

In our analysis, we employed Xspec v12.9.1~\citep{arnaud}. We fitted the data with five models, which are briefly described below. For each model, we allow one of the Johannsen deformation parameters (either $\alpha_{13}$ or $\alpha_{22}$) to vary at a time while keeping the other one fixed at 0.

While our goal is to test general relativity, the comparison of different models, from simple to sophisticated, is necessary to find the correct one. Note that we cannot rely on previous studies that assumed the Kerr metric: we may indeed find that a deviation from the Kerr spacetime could explain the data without adding some component that seems instead necessary when we assume the Kerr metric. We do compare our results with previous results, but only {\it a posteriori}.

\subsection*{Model~$a$}

Model~$a$ is the simplest model and includes just a power-law
\be
\text{\sc tbabs*zpowerlaw} \, . \nonumber
\ee
{\sc tbabs} describes the galactic absorption~\citep{wilms} and we fixed the galactic column density to $N_{\rm H} = 6.74 \cdot 10^{20}$~cm$^{-2}$~\cite{nH,nH2}. {\sc zpowerlaw} describes a redshifted photon power-law spectrum. This is the simplest model and only describes the spectrum of the corona. As we can see from panels~$(a)$ in Fig.~\ref{f-ratio}, which show the data to best-fit model ratio, we have an excess of photon count at low energies and a broad iron line around 6.4~keV. This suggests to us that we have to add a relativistically blurred reflection component. The best-fit values are reported in the second column in Tab.~\ref{t-fit} and in Tab.~\ref{t-fit2}. Since this model does not include the Johannsen deformation parameters, Tabs.~\ref{t-fit} and \ref{t-fit2} show the same result.

\subsection*{Model~$b$}

This is the next-to-simplest model and adds the reflection spectrum from the accretion disk
\be
\text{\sc tbabs*relxill\_nk} \, . \nonumber
\ee
{\sc relxill\_nk} would describe both the power-law component and the disk's reflection spectrum, but we find that, once we have added the reflection spectrum, the power-law component is not necessary and therefore we set the reflection fraction parameter $R = -1$ (no power-law component). This can be explained with a corona extremely close to the black hole, so that most of the radiation from the corona is bent toward the disk and the fraction capable of escaping to infinity is very small. The data to best-fit model ratios are shown in panels~$(b)$ in Fig.~\ref{f-ratio}, and the best-fit values are reported in the third column in Tab.~\ref{t-fit} ($\alpha_{13}$ free and $\alpha_{22} = 0$) and Tab.~\ref{t-fit2} ($\alpha_{13} = 0$ and $\alpha_{22}$ free).

\subsection*{Model~$c$} 

We consider a double reflection model, which is currently quite a popular choice to fit the X-ray spectrum of some narrow line Seyfert~1 galaxies
\be
\text{\sc tbabs*(relxill\_nk + relxill\_nk)} \, . \nonumber
\ee
The general idea behind this model is that there are certain inhomogeneities in the accretion disk, but the origin and nature of these inhomogeneities can vary. For instance, the density of the disk photosphere may be patchy, leading to mixed regions of high and low ionization~\cite{lohfink}; the surface of the disk may have regions of different density~\cite{fabian11}; it is possible that we are looking at a disk with different layers~\cite{kara15}. Note that the parameters of the two reflection components are tied with the exception of the ionization $\xi$ and the normalization. As we can see from panels~$(c)$ in Fig.~\ref{f-ratio} and the fourth column in Tab.~\ref{t-fit} ($\alpha_{13}$ free and $\alpha_{22} = 0$) and in Tab.~\ref{t-fit2} ($\alpha_{13} = 0$ and $\alpha_{22}$ free), adding a second reflection component only leads to a very modest improvement.

\subsection*{Model~$d$}

We consider the model
\be
\text{\sc tbabs*(relxill\_nk + xillver)} \, . \nonumber
\ee
Here we have two reflectors, namely the accretion disk and some non-relativistic material. It is indeed quite natural to expect the presence of gas around supermassive black holes and not directly participating in the accretion process. {\sc relxill\_nk} describes the disk's reflection spectrum. {\sc xillver} describes the reflection spectrum from such a warm material at a larger distance from the black hole and is independent of the background metric~\cite{r3}. We assume the same iron abundance $A_{\rm Fe}$ in {\sc relxill\_nk} and in {\sc xillver}, while the ionization is independent. As we can see from panels~$(d)$ in Fig.~\ref{f-ratio}, the fit is better than models~$(b)$ and $(c)$. The best-fit values are reported in the fifth column in Tab.~\ref{t-fit} and in Tab.~\ref{t-fit2}.

\subsection*{Model~$e$}

\begin{figure*}[t]
\begin{center}
\includegraphics[type=pdf,ext=.pdf,read=.pdf,width=8.5cm]{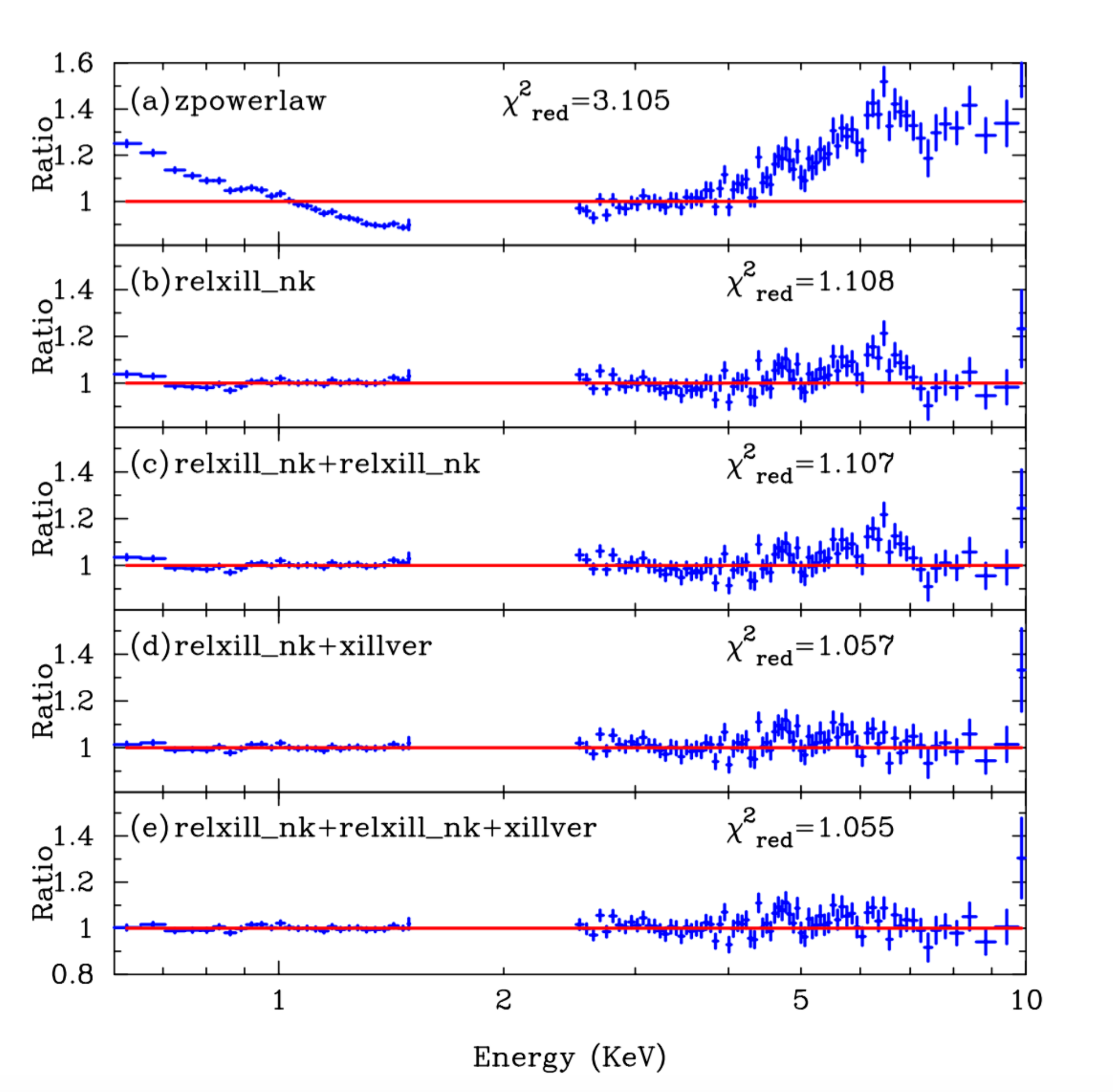}
\includegraphics[type=pdf,ext=.pdf,read=.pdf,width=8.5cm]{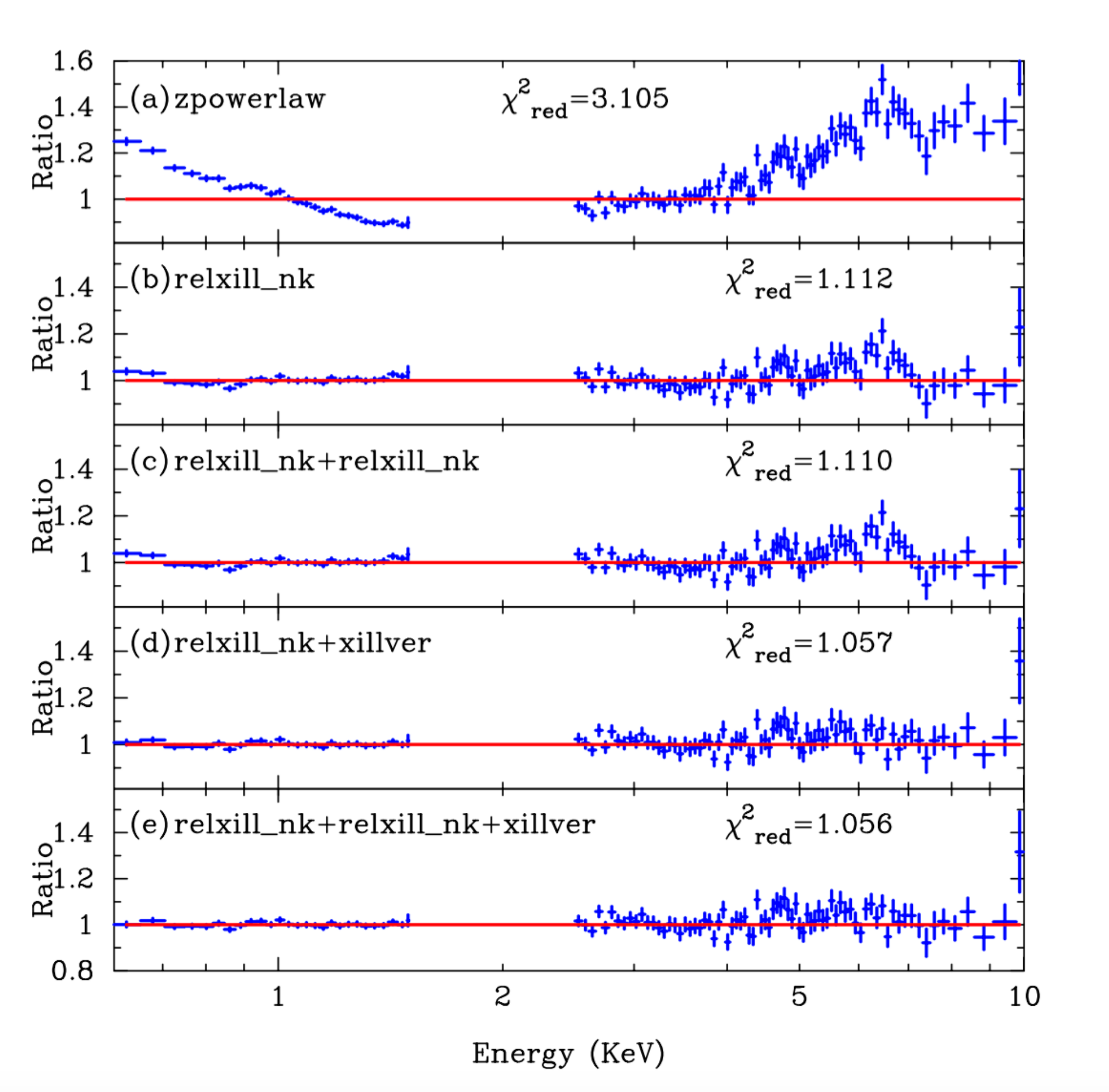}
\end{center}
\vspace{-0.3cm}
\caption{Data to best-fit model ratios for the spectral models $a$ to $e$. In the left panel, $\alpha_{13}$ is free in the fit and $\alpha_{22}=0$. In the right panel, $\alpha_{13}=0$ and $\alpha_{22}$ can vary. See the text for more details. \label{f-ratio}}
\vspace{0.0cm}
\begin{center}
\hspace{-0.5cm}
\includegraphics[type=pdf,ext=.pdf,read=.pdf,width=9.0cm]{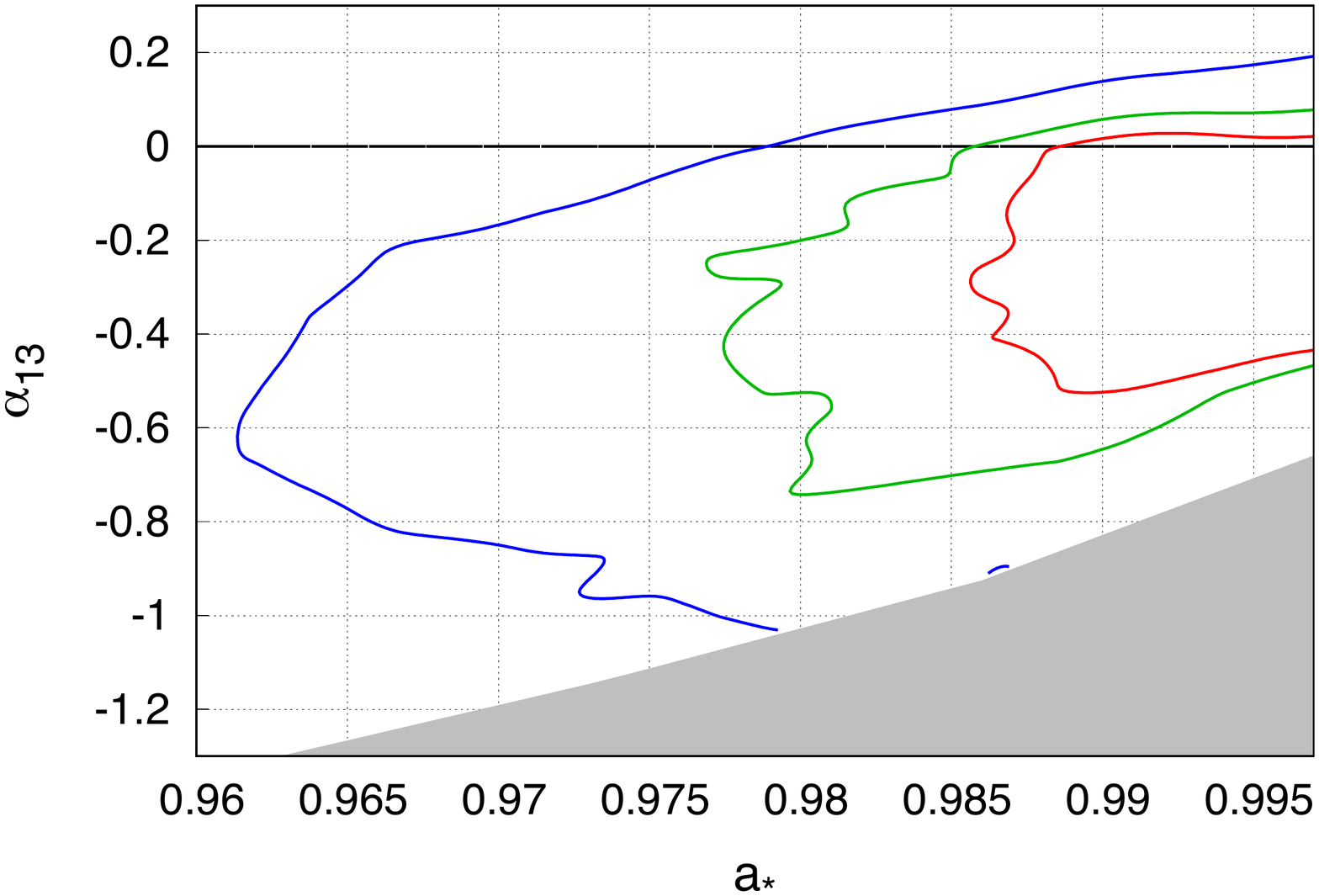}
\includegraphics[type=pdf,ext=.pdf,read=.pdf,width=9.0cm]{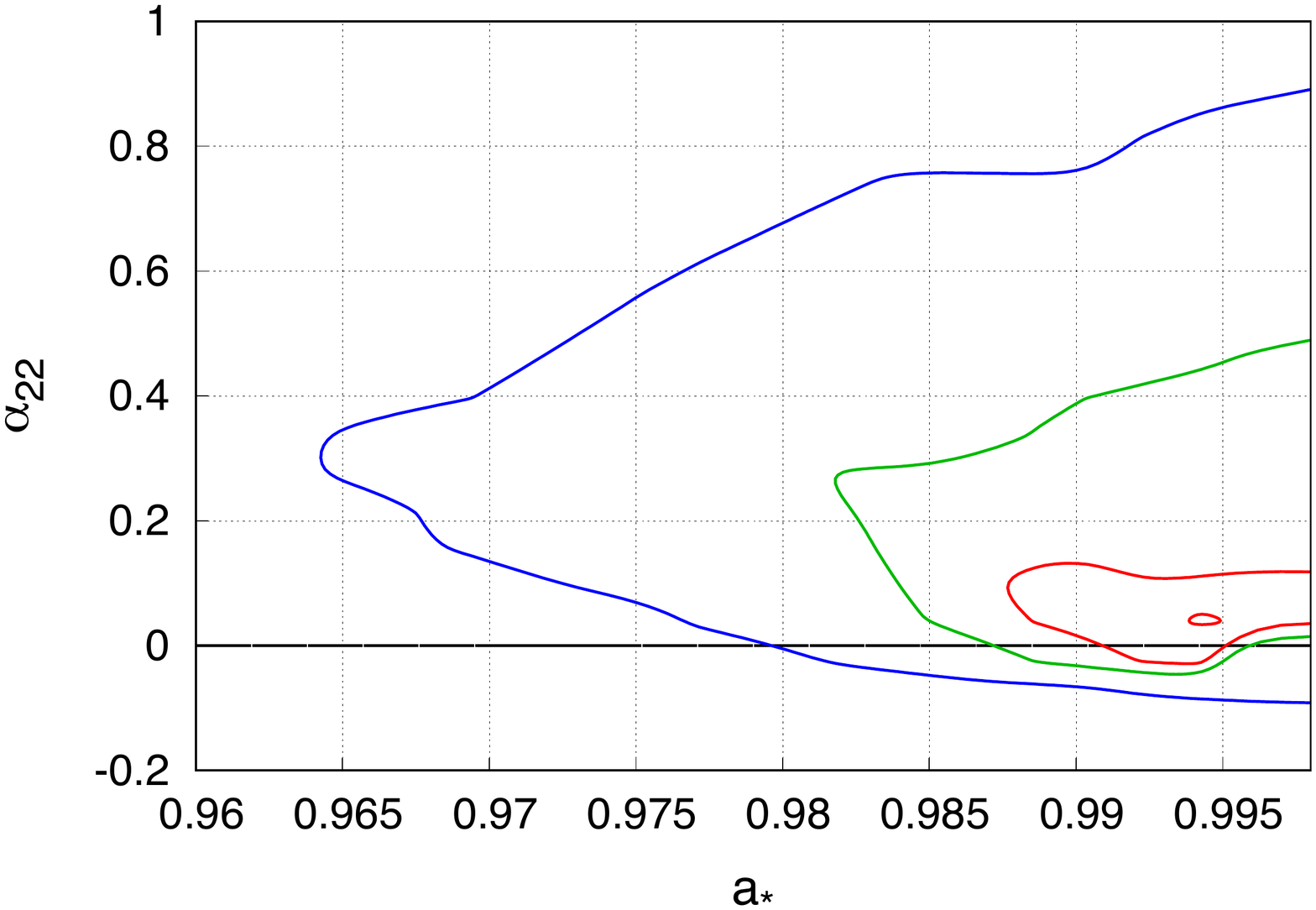}
\end{center}
\vspace{-1.2cm}
\caption{Constraints on the spin parameter $a_*$ and the Johannsen deformation parameters $\alpha_{13}$ (left panel) and $\alpha_{22}$ (right panel) from the \textsl{Suzaku} data of the supermassive black hole in Ark~564 assuming model~$d$. The red, green, and blue lines indicate, respectively, the 68\%, 90\%, and 99\% confidence level contours for two relevant parameters. The gray region in the left panel is excluded in our analysis because the metric is singular there. See the text for more details. \label{f-m4}}
\end{figure*}

Last, we consider a double reflection model for the accretion disk and a warm non-relativistic material
\be
\text{\sc tbabs*(relxill\_nk + relxill\_nk + xillver)} \, . \nonumber
\ee
In the three reflection components, the iron abundance is the same, while the ionization parameters are all independent. The data to best-fit model ratios are shown in panels~$(e)$ in Fig.~\ref{f-ratio} and the best-fit values are reported in the sixth column in Tab.~\ref{t-fit} and in Tab.~\ref{t-fit2}. As we could have expected on the basis of previous results, the improvement with respect to model~$d$ is modest: the data do not seem to require a double reflection spectrum.

\begin{table*}
\centering
\vspace{0.8cm}
\begin{tabular}{lcccccc}
\hline\hline
Model & $a$ & $b$ & $c$ & $d$ & $e$ \\
\hline
{\sc tbabs} &&&& \\
$N_{\rm H} / 10^{20}$ cm$^{-2}$ & 6.74$^\star$ & 6.74$^\star$ & 6.74$^\star$ & 6.74$^\star$ & 6.74$^\star$ \\
\hline
{\sc zpowerlaw} &&&&& \\
$\Gamma$ & $2.768^{+0.003}_{-0.003}$ & -- & -- & -- & -- \\
$z$ & 0.0247$^\star$ & -- & -- & -- & -- \\
\hline
{\sc relxill\_nk} &&&&& \\
$q$ & -- & $> 8.4$ & $9.86_{-1.1}^{+0.08}$ & $9.62_{-0.74}^{+0.16}$ & $7.0_{-0.5}^{+1.6}$ \\
$i$ [deg] & -- & $26.7_{-2.7}^{+2.9}$ & $26.6_{-5.3}^{+2.6}$ & $36.0_{-0.7}^{+4.3}$ & $<45$ \\
$a_*$ & -- & $>0.987$ & $0.993_{-0.012}^{+0.004}$ & $0.995_{-0.007}^{+0.001}$ & $0.988_{-0.010}^{+0.003}$ \\
$\alpha_{22}$ & -- & $-0.1^{+0.1}_{-0.1}$ & $-0.1^{+0.1}_{-0.1}$ & $0_{-0.05}^{+0.3}$ & $-0.1_{-0.1}$ \\
$z$ & -- & 0.0247$^\star$ & 0.0247$^\star$ & 0.0247$^\star$ & 0.0247$^\star$ \\
$\log\xi$ & -- & $3.088_{-0.011}^{+0.011}$ & $4.4_{-0.0}^{+0.3}$ & $2.66_{-0.04}^{+0.06}$ & $2.72_{-0.23}^{+0.10}$ \\
$A_{\rm Fe}$ & -- & $0.525_{-0.025}^{+0.0}$ & $0.53_{-0.03}^{+0.0}$ & $0.88_{-0.06}^{+0.12}$ & $0.70_{-0.20}^{+0.0}$ \\
$R$ & -- & $-1$ & $-1$ & $-1$ & $-1$ \\
\hline
{\sc relxill\_nk} &&&&& \\
$\log\xi'$ & -- & -- & $3.01_{-0.06}^{+0.05}$ & -- & $1.3_{-1.3}^{+0.2}$ \\
\hline
{\sc xillver} &&&&& \\
$\log\xi''$ & -- & -- & -- & $4.42_{-0.22}^{+0.0}$ & $4.46_{-0.0}^{+0.24}$ \\ 
\hline\hline
$\chi^2$/dof & \hspace{0.2cm} 4356.16/1403 \hspace{0.2cm} & \hspace{0.2cm} 1553.01/1397 \hspace{0.2cm} & \hspace{0.2cm} 1548.65/1395 \hspace{0.2cm} & \hspace{0.2cm} 1474.58/1395 \hspace{0.2cm} & \hspace{0.2cm} 1471.61/1393 \hspace{0.2cm} \\
& = 3.105 & = 1.112 & = 1.110 & = 1.057 & = 1.056 \\
\hline\hline
\end{tabular}
\caption{As in Tab.~\ref{t-fit} assuming $\alpha_{13}=0$ and $\alpha_{22}$ free in the fits. Model~$a$ is the same as in Tab.~\ref{t-fit} because it does not include {\sc relxill\_nk}, but we report here for completeness. \label{t-fit2}}
\end{table*}

\section{Results\label{sec:results}}
In all models with {\sc relxill\_nk}, we find a high value of the photon index $q$; that is, most of the radiation seems to come from the very inner part of the accretion disk. The spin parameter $a_*$ is always very close to 1. This was found even in Ref.~\cite{walton} assuming a Kerr background. Our best-fit values of $a_*$ are somewhat higher than the result in Ref.~\cite{walton}, consistent with the fact that the {\sc relxill} package finds higher spin values than {\sc reflionx}, which is the reflection model employed in~\cite{walton}. The inclination angle of the disk with respect to our line of sight, $i$, is not high but cannot be constrained well. The iron abundance $A_{\rm Fe}$ (in units of Solar iron abundance) is always less than 1. Note that, in models~$d$ and $e$, the non-relativistic component described by {\sc xillver} is subdominant with respect to the relativistic one described by {\sc relxill\_nk}. This permits us to get measurements of $a_*$ and of the deformation parameters.

The main goal of our study is to constrain the spacetime metric around the supermassive black hole in Ark~564, getting a measurement of $a_*$ $\alpha_{13}$, and $\alpha_{22}$. Models~$d$ and $e$ provide equally good fit, so we choose model~$d$ as our best model because it is simpler and the second reflection spectrum looks unnecessary (Occam's razor). The constraints on $a_*$, $\alpha_{13}$, and $\alpha_{22}$ from model~$d$ are shown in Fig.~\ref{f-m4}. In the left panel, we have the constraints on $a_*$ and $\alpha_{13}$ assuming $\alpha_{22}=0$. In the right panel we have the opposite case with the constraints on $a_*$ and $\alpha_{22}$ for $\alpha_{13}=0$. The red, green, and blue lines indicate, respectively, the 68\%, 90\%, and 99\% confidence level contours for two relevant parameters. The black solid lines at $\alpha_{13} = 0$ and $\alpha_{22} = 0$ mark the Kerr solution. The gray region is excluded from analysis since it violates the constraint of Sec.~\ref{sec:metric}.

Our results are consistent with the hypothesis that the supermassive object in Ark~564 is a Kerr black hole. Assuming $\alpha_{22} = 0$, the constraints on $a_*$ and $\alpha_{13}$ are (at 99\%~confidence level)  
\be\label{eq-a13-w}
a_* > 0.96 \, , \quad
-1.0 < \alpha_{13} < 0.2 \, .
\ee
For the case $\alpha_{13} = 0$, we find the following constraints on $a_*$ and $\alpha_{22}$ (at 99\%~confidence level)
\be
a_* > 0.96 \, , \quad
-0.1 < \alpha_{22} < 0.9 \, .
\ee
The constraints on $\alpha_{13}$ in Eq.~(\ref{eq-a13-w}) can be compared to those obtained previously in~\cite{noi2} (see also Ref.~\cite{universe}) from the supermassive black hole in 1H0707--495. The constraint reported here is in agreement with and stronger than the constraint from 1H0707--495.

We checked the significance of adding the deformation parameters with an F-test. Model~$d$ with $\alpha_{13}$ or $\alpha_{22}$ has $\chi^2 \approx 1474$ (with a small difference between the two cases) and the degrees of freedom are 1395. The restricted hypothesis ($\alpha_{13} = \alpha_{22} = 0$) has $\chi^2 \approx 1476$ and the degrees of freedom are 1396. The models with non-zero deformation parameters are comparable to the Kerr one at a confidence level greater than 85\%.

\section{Discussion\label{sec:discuss}}

We would like to remark that, even if our analysis nicely confirms the Kerr black hole hypothesis, these constraints have to be taken with some caution. Our error only includes the statistical uncertainty, ignoring the systematic uncertainty. Our model, like any astrophysical model, has a number of simplifications which inevitably lead to systematic uncertainties in the analysis. Here we discuss some aspects of these systematic uncertainties.

First, let us note that the models labeled $b$ and $c$ were a poor fit the data (as can be seen from their ratio plots in Fig.~\ref{f-ratio}). Therefore the fact that the associated measurements of the deformation parameters is not consistent with the Kerr hypothesis (see Tab.~\ref{t-fit} and \ref{t-fit2}) is irrelevant, because both models are clearly wrong. Model~$d$ could fit well the data and turned out to require the Kerr metric. This shows that getting the correct combination of models to describe the data is critical for testing the metric.

We also note that Ark~564 is quite a variable source and the disk and corona geometries likely change with time. From the \textsl{Suzaku} data analyzed in the present paper, we find that the inner edge of the disk is extremely close to the black hole, and we do not see any significant power-law component, suggesting that the corona is also very close to the black hole. In Ref.~\cite{kara17}, the authors, when using data from \textsl{NuSTAR}, find a steep spectrum with a low cut-off energy for the corona, with no relativistic reflection; whereas when using data from both \textsl{NuSTAR} and \textsl{Suzaku}, they find that relativistic reflection is indeed required to fit the data correctly. They conclude that the correct model could be a combination of the two. This matches perfectly with our findings here. Using only \textsl{Suzaku} data, we find that a combination of relativistic and ionized reflection models best describes the data.

Additional assumptions that go into the model have to be accounted for in future. For instance, the accretion disk is assumed to be infinitesimally thin and the inner edge of the disk is set at the ISCO radius. The intensity profile is modeled with a simple power-law, which is surely an approximation. The accretion disk has a unique ionization parameter. We allow only one deformation parameter to vary at a time, whereas in principle many of them may simultaneously be non-zero. We plan to improve our reflection model and study the systematic uncertainties in future works.

\vspace{0.4cm}


{\bf Acknowledgments --}
We thank Matteo Guainazzi for reading a preliminary version of this manuscript and providing useful feedback. This work was supported by the National Natural Science Foundation of China (NSFC), Grant No.~U1531117, and Fudan University, Grant No.~IDH1512060. A.T. also acknowledges support from the China Scholarship Council (CSC), Grant No.~2016GXZR89. S.N. acknowledges support from the Excellence Initiative at Eberhard-Karls Universit\"at T\"ubingen. A.B.A. also acknowledges the support from the Shanghai Government Scholarship (SGS). J.J. is supported by the Cambridge Trust and the Chinese Scholarship Council Joint Scholarship Programme (201604100032). C.B. also acknowledges support from the Alexander von Humboldt Foundation.



\begin{thebibliography}{99}

\bibitem{will} 
  C.~M.~Will,
  Living Rev.\ Rel.\  {\bf 17}, 4 (2014)
  [arXiv:1403.7377 [gr-qc]].
  
\bibitem{rev} 
  C.~Bambi,
  Rev.\ Mod.\ Phys.\  {\bf 89}, 025001 (2017)
  [arXiv:1509.03884 [gr-qc]].    
  
\bibitem{kerr1} 
  B.~Carter,
  Phys.\ Rev.\ Lett.\  {\bf 26}, 331 (1971).
  
\bibitem{kerr2} 
  D.~C.~Robinson,
  Phys.\ Rev.\ Lett.\  {\bf 34}, 905 (1975).    
  
\bibitem[Bambi(2017b)]{book} 
  C.~Bambi,
  {\it Black Holes: A Laboratory for Testing Strong Gravity} (Springer Singapore, 2017).  
  
\bibitem{new} 
  E.~Berti {\it et al.},
  Class.\ Quant.\ Grav.\  {\bf 32}, 243001 (2015)
  [arXiv:1501.07274 [gr-qc]].
  
\bibitem{n1} 
  S.~D.~Mathur,
  Fortsch.\ Phys.\  {\bf 53}, 793 (2005)
  [hep-th/0502050].  
  
\bibitem{n2} 
  G.~Dvali and C.~Gomez,
  Fortsch.\ Phys.\  {\bf 61}, 742 (2013)
  [arXiv:1112.3359 [hep-th]].  
  
\bibitem{n3} 
  S.~B.~Giddings,
  Phys.\ Rev.\ D {\bf 90}, 124033 (2014)
  [arXiv:1406.7001 [hep-th]].

\bibitem{n4} 
  S.~B.~Giddings,
  Nature Astronomy {\bf 1}, 0067 (2017)
  [arXiv:1703.03387 [gr-qc]].
  
\bibitem{k0} 
  A.~C.~Fabian, K.~Iwasawa, C.~S.~Reynolds and A.~J.~Young,
  Publ.\ Astron.\ Soc.\ Pac.\  {\bf 112}, 1145 (2000)
  [astro-ph/0004366].  
  
\bibitem{k1} 
  C.~S.~Reynolds,
  Space Sci.\ Rev.\  {\bf 183}, 277 (2014)
  [arXiv:1302.3260 [astro-ph.HE]].

\bibitem{k2} 
  M.~C.~Miller and J.~M.~Miller,
  Phys.\ Rept.\  {\bf 548}, 1 (2015)
  [arXiv:1408.4145 [astro-ph.HE]].
  
\bibitem{nk0} 
  J.~Schee and Z.~Stuchlik,
  Gen.\ Rel.\ Grav.\  {\bf 41}, 1795 (2009)
  [arXiv:0812.3017 [astro-ph]].
    
\bibitem{nk1} 
  T.~Johannsen and D.~Psaltis,
  Astrophys.\ J.\  {\bf 773}, 57 (2013)
  [arXiv:1202.6069 [astro-ph.HE]].  
  
\bibitem{nk2} 
  C.~Bambi,
  Phys.\ Rev.\ D {\bf 87}, 023007 (2013)
  [arXiv:1211.2513 [gr-qc]].
  
\bibitem{nk2a} 
  J.~Jiang, C.~Bambi and J.~F.~Steiner,
  JCAP {\bf 1505}, 025 (2015)
  [arXiv:1406.5677 [gr-qc]].  
  
\bibitem{nk2b} 
  J.~Jiang, C.~Bambi and J.~F.~Steiner,
  Astrophys.\ J.\  {\bf 811}, 130 (2015)
  [arXiv:1504.01970 [gr-qc]].  
  
\bibitem{nk3} 
  C.~Bambi, J.~Jiang and J.~F.~Steiner,
  Class.\ Quant.\ Grav.\  {\bf 33}, 064001 (2016)
  [arXiv:1511.07587 [gr-qc]].  
    
\bibitem{nk3b} 
  M.~Zhou, A.~Cardenas-Avendano, C.~Bambi, B.~Kleihaus and J.~Kunz,
  Phys.\ Rev.\ D {\bf 94}, 024036 (2016)
  [arXiv:1603.07448 [gr-qc]].    
    
\bibitem{nk4} 
  Y.~Ni, M.~Zhou, A.~Cardenas-Avendano, C.~Bambi, C.~A.~R.~Herdeiro and E.~Radu,
  JCAP {\bf 1607}, 049 (2016)
  [arXiv:1606.04654 [gr-qc]].  
  
\bibitem{nk5} 
  S.~Nampalliwar, C.~Bambi, K.~Kokkotas and R.~Konoplya,
  Phys.\ Lett.\ B {\bf 781}, 626 (2018)
  [arXiv:1803.10819 [gr-qc]].   
    
\bibitem{r1} 
  T.~Dauser, J.~Garcia, J.~Wilms, M.~Bock, L.~W.~Brenneman, M.~Falanga, K.~Fukumura and C.~S.~Reynolds,
  Mon.\ Not.\ Roy.\ Astron.\ Soc.\  {\bf 430}, 1694 (2013)
  [arXiv:1301.4922 [astro-ph.HE]].

\bibitem{r2} 
  J.~Garc{\'\i}a {\it et al.},
  Astrophys.\ J.\  {\bf 782}, 76 (2014)
  [arXiv:1312.3231 [astro-ph.HE]].

\bibitem{noi1} 
  C.~Bambi, A.~Cardenas-Avendano, T.~Dauser, J.~A.~Garcia and S.~Nampalliwar,
  Astrophys.\ J.\  {\bf 842}, 76 (2017)
  [arXiv:1607.00596 [gr-qc]].

\bibitem{noi2} 
  Z.~Cao, S.~Nampalliwar, C.~Bambi, T.~Dauser and J.~A.~Garcia,
  Phys.\ Rev.\ Lett.\  {\bf 120}, 051101 (2018)
  [arXiv:1709.00219 [gr-qc]].  
  
\bibitem{fabian11} 
  A.~C.~Fabian {\it et al.},
  Mon.\ Not.\ Roy.\ Astron.\ Soc.\  {\bf 419}, 116 (2012)
  [arXiv:1108.5988 [astro-ph.HE]].
  
\bibitem{kara15} 
  E.~Kara {\it et al.},
  Mon.\ Not.\ Roy.\ Astron.\ Soc.\  {\bf 449}, 234 (2015)
  [arXiv:1501.06849 [astro-ph.HE]].  
  
\bibitem{Wang-Ji:2018ssh} 
  J.~Wang-Ji, A.~B.~Abdikamalov, D.~Ayzenberg, C.~Bambi, T.~Dauser, J.~A.~Garcia, S.~Nampalliwar and J.~F.~Steiner,
  arXiv:1806.00126 [gr-qc].  
  
\bibitem{j-m} 
  T.~Johannsen,
  Phys.\ Rev.\ D {\bf 88}, 044002 (2013)
  [arXiv:1501.02809 [gr-qc]].    
  
\bibitem{walton} 
  D.~J.~Walton, E.~Nardini, A.~C.~Fabian, L.~C.~Gallo and R.~C.~Reis,
  Mon.\ Not.\ Roy.\ Astron.\ Soc.\  {\bf 428}, 2901 (2013)
  [arXiv:1210.4593 [astro-ph.HE]]. 

\bibitem{kara13} 
  E.~Kara, A.~C.~Fabian, E.~M.~Cackett, P.~Uttley, D.~R.~Wilkins and A.~Zoghbi,
  Mon.\ Not.\ Roy.\ Astron.\ Soc.\  {\bf 434}, 1129 (2013)
  [arXiv:1306.2551 [astro-ph.HE]].
  
\bibitem{giustini} 
  M.~Giustini, T.~J.~Turner, J.~N.~Reeves, L.~Miller, E.~Legg, S.~B.~Kraemer and I.~M.~George,
  Astron.\ Astrophys.\  {\bf 577}, A8 (2015)
  [arXiv:1502.01338 [astro-ph.HE]].

\bibitem{kara17} 
  E.~Kara, J.~A.~Garcia, A.~Lohfink, A.~C.~Fabian, C.~S.~Reynolds, F.~Tombesi and D.~R.~Wilkins,
  Mon.\ Not.\ Roy.\ Astron.\ Soc.\  {\bf 468}, 3489 (2017)
  [arXiv:1703.09815 [astro-ph.HE]].
  
\bibitem{koyama} 
  K.~Koyama {\it et al.},
  Publ.\ Astron.\ Soc.\ Jap.\  {\bf 59}, S23 (2007).  
        
\bibitem{arnaud} 
  K.~A.~Arnaud,
  Astronomical Data Analysis Software and Systems V, {\bf 101}, 17 (1996).

\bibitem{wilms} 
  J.~Wilms, A.~Allen and R.~McCray,
  Astrophys.\ J.\  {\bf 542}, 914 (2000)
  [astro-ph/0008425].

\bibitem{nH} 
  http://www.swift.ac.uk/analysis/nhtot/
  
\bibitem{nH2}
  R.~Willingale, R.~L.~C.~Starling, A.~P.~Beardmore, N.~R.~Tanvir and P.~T.~O'Brien,
  Mon.\ Not.\ Roy.\ Astron.\ Soc.\  {\bf 431}, 394 (2013)
  [arXiv:1303.0843 [astro-ph.HE]].     

\bibitem{lohfink} 
  A.~M.~Lohfink, C.~S.~Reynolds, J.~M.~Miller, L.~W.~Brenneman, R.~F.~Mushotzky, M.~A.~Nowak and A.~C.~Fabian,
  Astrophys.\ J.\  {\bf 758}, 67 (2012)
  [arXiv:1209.0468 [astro-ph.HE]].  

\bibitem{r3} 
  J.~Garcia, T.~Dauser, C.~S.~Reynolds, T.~R.~Kallman, J.~E.~McClintock, J.~Wilms and W.~Eikmann,
  Astrophys.\ J.\  {\bf 768}, 146 (2013)
  [arXiv:1303.2112 [astro-ph.HE]].
  
\bibitem{universe} 
  C.~Bambi {\it et al.},
  Universe {\bf 4}, 79 (2018)
  [arXiv:1806.02141 [gr-qc]].  

\end{thebibliography}
\end{document}